\documentclass[letterpaper,12pt]{article}

\usepackage{multicol}
\usepackage{multirow}
\usepackage{subfig}
\usepackage{caption}
\usepackage{amsmath}
\usepackage{setspace}
\usepackage{natbib}
\usepackage{enumitem}

\usepackage{fullpage}
\usepackage[dvips]{graphicx}

\usepackage[usenames]{color}

\usepackage{mathtools}

\usepackage{courier}

\usepackage{upgreek}

\singlespace
\bibpunct{(}{)}{;}{a}{,}{,}

\begin{document}

\noindent {\bf TITLE:} Parameter estimation for the generalized extreme value distribution: a method that combines bootstrapping and $r$ largest order statistics
\\
\\
{\bf AUTHOR:} Juan L.P. Soto, Department of Telecommunications and Control Engineering, University of S\~{a}o Paulo, S\~{a}o Paulo, Brazil
\\
\\
{\bf CORRESPONDING AUTHOR:} Juan L.P. Soto, Department of Telecommunications and Control Engineering, University of S\~{a}o Paulo, Av. Prof. Luciano Gualberto, tr. 3, n. 158, 05508-900, S\~{a}o Paulo, Brazil, email: juansoto@usp.br 
\\

\begin{center} {\bf \large Abstract} \end{center}

A critical problem in extreme value theory (EVT) is the estimation of parameters for the limit probability distributions. Block maxima (BM), an approach in EVT that seeks estimates of parameters of the generalized extreme value distribution (GEV), can be generalized to take into account not just the maximum realization from a given dataset, but the $r$ largest order statistics for a given $r$. In this work we propose a parameter estimation method that combines the $r$ largest order statistic ($r$-LOS) extension of BM with permutation bootstrapping: surrogate realizations are obtained by randomly reordering the original data set, and then $r$-LOS is applied to these shuffled measurements — the mean estimate computed from these surrogate realizations is the desired estimate. We used synthetic observations and real meteorological time series to verify the performance of our method; we found that the combination of $r$-LOS and bootstrapping resulted in estimates more accurate than when either approach was implemented separately.





\section{Introduction}

Extreme value theory (EVT) \citep{Coles_2001}, the field in statistical analysis that investigates the behavior of maxima of random variables, has been attracting increasing interest from a diverse range of scientific communities in the past few decades: recent applications of EVT include traffic safety \citep{Wang_2019, Borsos_2020, Ali_2023}, fault detection in mechanical systems \citep{Toshkova_2020, Yu_2021, Wang_2023}, medicine \citep{Schipaanboord_2018, Szigeti_2023}, meteorology \citep{Ban_2020}, oceanography \citep{Tendijck_2023}, finance \citep{Liu_2018, MartinsFilho_2018, Novales_2019}, astrophysics \citep{Acero_2018, Elvidge_2020}, athlete performance \citep{Spearing_2021}, maritime radar \citep{Shui_2022} and cryptocurrencies \citep{Gkillas_2018}. Usually, investigations involving EVT are based on limit probability distributions whose parameters can be estimated from a sequence of realizations of the random variable of interest. The two main approaches are: 
\begin{itemize}
\item block maxima (BM), where the realizations are separated into equal-sized blocks, the maximum realization of each block is determined, and from these maxima the estimates of the parameters of the generalized extreme value distribution (GEV) are calculated; and 
\item peaks over threshold (POT), where the parameters of the generalized Pareto distribution (GPD) are calculated from those observations that are higher than a pre-specified value. 
\end{itemize}     
One extension of BM is the $r$ largest order statistics approach ($r$-LOS) \citep{Smith_1986, Coles_2001, Zhang_2004}, which uses for the estimation of the GEV parameters, instead of only the maximum realization within each block, the $r$ highest values of the block, for a given $r$; as more information is extracted from the data for the subsequent analyses (thus resulting in potentially more precise estimates), $r$-LOS is a popular alternative to BM in a variety of studies \citep{An_2007, Wang_2008, Feng_2015, Naseef_2017, Sikhwari_2022, Haltas_2022}.  

In this work, we propose a modification of the $r$-LOS approach based on a variant of the bootstrapping framework \citep{Efron_1979}. \cite{Mefleh_2021} presented a procedure, {\it permutation bootstrapping,} that sought to increase the accuracy of parameter estimates computed with BM by
\begin{enumerate}[label=(\alph*)]
\item randomly shuffling the order of the sequence of observations;
\item applying BM to this surrogate collection of realizations;
\item repeating steps (a) and (b) several times to obtain a set of surrogate estimates;
\item compute the mean of these estimates to find the bootstrap estimates.
\end{enumerate}
Our goal is to combine the improved performance of permutation bootstrapping (when compared with standard BM), as demonstrated both theoretically and empirically by \cite{Mefleh_2021}, with the potential of $r$-LOS for more precise estimations (given the use of more data from each block) so that the resulting approach yields more accurate estimates than either $r$-LOS or BM implemented separately. The evaluation of all methods considered here is carried out with numerical simulations and with real meteorological time series.

\section{Methods}
\label{sec_methods}

All computations were performed with R, and the plots were created with MATLAB; the scripts used to run these calculations are available from the author upon request. Some analyses made use of the packages \texttt{pareto}, \texttt{ismev} and \texttt{invgamma}. 

\subsection{Extreme value theory: block maxima}
\label{sec_evt}

Let $X_1,...,X_n$ be a sequence of independent random variables that follow a common probability distribution $F$, and $M_n = \max \{ X_1,...,X_n \}$. The Fisher-Tippett-Gnedenko theorem states that, if there exist sequences of normalizing scalars $a_n > 0$ and $b_n$ such that the probability distribution $P[(M_n-b_n)/a_n \leq y_n]$ converges in distribution to a non-degenerate function $G(y)$, then the latter follows a GEV distribution \citep{Coles_2001, Mefleh_2021}:
\begin{equation}
G(y) = \left\{
\begin{array}{lll}
\exp \left\{ -\left[   1 + \xi \left( \dfrac{y-\mu}{\sigma} \right)  \right]^{-1/\xi} \right\}, & \xi \neq 0, &  1+\xi (y-\mu)/\sigma > 0; \\
& & \\
\exp \left[ -\exp \left( -\dfrac{y-\mu}{\sigma} \right) \right], & \xi = 0, &
\end{array}
\right.
\label{eq_gev}
\end{equation}
where $\mu$ and $\sigma > 0$ are the location and scale parameters, respectively, and $\xi$ is the extreme value index (or shape parameter). 
Equation (\ref{eq_gev}) unifies three families of extreme value distributions, according to the value of the shape parameter: $\xi > 0$ results in the Fréchet family of distributions, $\xi < 0$ becomes the Weibull family, and $\xi = 0$ yields the Gumbel family.  

We can obtain estimates of the GEV parameters $\mu$, $\sigma$ and $\xi$ of equation (\ref{eq_gev}), based on realizations $x_1,...,x_n$ of the sequence of random variables $X_1,...,X_n$, by means of the block maxima (BM) approach. Given a block of size $s$, let $y_1,...,y_m$ $(m \times s \leq n)$ be the block maxima, i.e. $y_1 = \max \{ x_1,...,x_s \}$, $y_2 = \max \{ x_{s+1},...,x_{2s} \}$ and so on. The log-likelihood for the GEV parameters is \citep{Coles_2001}
\begin{equation}
\ell(\mu,\sigma,\xi) = -m \log \sigma -\left(1+\frac{1}{\xi}\right) \sum^m_{i=1} \log \left[ 1+\xi \left( \frac{y_i-\mu}{\sigma} \right) \right] - \sum^m_{i=1} \left[ 1+\xi \left( \frac{y_i-\mu}{\sigma} \right) \right]^{-1/\xi} ,
\label{eq_like}
\end{equation}
when $\xi \neq 0$, and provided $1 + \xi (y_i-\mu)\sigma > 0$ for all $i$, and
\begin{equation}
\ell(\mu,\sigma) = -m\log \sigma - \sum^m_{i=1} \left( \frac{y_i-\mu}{\sigma} \right) - \sum^m_{i=1} \exp \left\{ - \left( \frac{y_i-\mu}{\sigma} \right) \right\} .
\label{eq_like_0}
\end{equation}
when $\xi = 0$. Though there is no analytical, closed-form solution to the equations above, numerical approximations can be obtained via optimization algorithms such as the Newton-Raphson method \citep{Hosking_1985, Macleod_1989}. Once adequate estimates of the GEV parameters $\hat\mu$, $\hat\sigma$ and $\hat\xi$ are found, estimates of extreme quantiles of 
$F$ can be computed by inverting equation (\ref{eq_gev}) and based on the relation $G = F^m$ \citep{Coles_2001, Mefleh_2021}:
\begin{equation}
\hat{q}_{F,p} = \hat{q}_{G,p^m} =
\left\{
\begin{array}{ll}
\hat\mu + (\hat\sigma/\hat\xi) \left[ \left( -m \log p \right)^{-\hat\xi} -1 \right] , & \text{if }  \hat\xi \neq 0;    \\
\hat\mu -\hat\sigma \log \left( -m \log p \right) , & \text{if }  \hat\xi = 0.
\end{array}
\right.
\label{eq_quant}
\end{equation}   

\subsection{$r$ largest order statistics}
\label{sec_rlos}

The method's description in this subsection is based on material presented by \cite{Coles_2001}.
A more general form of the Fisher-Tippett-Gnedenko theorem states that, if there exist sequences of normalizing scalars $a_n > 0$ and $b_n$ such that the probability distribution $P[(M_n-b_n)/a_n \leq y_n]$ converges in distribution to a non-degenerate function $G(y)$, then, for some fixed integer $r$, the limiting joint distribution for $n \to \infty$ of the random vector
$$ \tilde{\bf M}^{(r)}_n = \left( \dfrac{M^{(1)}_n-b_n}{a_n},\ldots,\dfrac{M^{(r)}_n-b_n}{a_n} \right), $$
where $M^{(k)}_n$ is the $k$ largest random variable in the sequence $\{ X_1,\ldots,X_n \}$, is associated with the joint density 
\begin{equation}
\begin{array}{rcl}
f\left(y^{(1)},\ldots,y^{(r)}\right) & = & \exp\left\{ - \left[ 1+\xi\left( \dfrac{y^{(r)}-\mu}{\sigma} \right) \right]^{-1/\xi} \right\} \\
& & \\
& & \times \displaystyle\prod^r_{k=1} \sigma^{-1} \exp \left[ 1+\xi\left( \dfrac{y^{(r)}-\mu}{\sigma} \right) \right]^{-1/\xi-1} ,
\end{array}
\end{equation}
where $\xi$ and $\mu$ are real-valued, $\sigma > 0$, $y^{(1)} \geq \ldots \geq y^{(r)}$, and the $y^{(k)}$ are such that $1+\xi(y^{(k)}-\mu)/\sigma > 0$ for all $k \in \{1,\ldots,r\}$. If $\xi = 0$, the density is 
\begin{equation}
f\left(y^{(1)},\ldots,y^{(r)}\right) = \exp\left\{ \left[ -\left( \dfrac{y^{(r)}-\mu}{\sigma} \right) \right] \right\} \times \prod^r_{k=1} \sigma^{-1} \exp \left[ -\left( \dfrac{y^{(k)}-\mu}{\sigma} \right) \right] .
\end{equation}
As with BM, we can obtain the GEV parameters by means of maximum likelihood estimation. Let $x_1,...,x_n$ be the realizations of $X_1,...,X_n$, which are then divided into $m$ blocks of size $s$. Further, let $y_i^{(k)}$ be the $k$ largest realization within block $i$, $i \in \{ 1,\ldots,m \}$, $k \in \{ 1,\ldots,r \}$, for some pre-defined $r$. The log-likelihood for the $r$-LOS approach is then
\begin{equation}
\begin{array}{rcl} 
\ell(\mu,\sigma,\xi) & = & mr \log \sigma - \displaystyle\sum^m_{i=1} \left[ 1+ \xi \left( \dfrac{y_i^{(r)}-\mu}{\sigma} \right) \right]^{-1/\xi}  \\
& & \\
& & - \displaystyle\sum^m_{i=1}\sum^r_{k=1}\left( \dfrac{1}{\xi} + 1 \right) \log \left[ 1 + \xi \left( \dfrac{y_i^{(k)}-\mu}{\sigma} \right) \right],
\end{array}
\label{eq_rlos_like}
\end{equation}    
if $\xi \neq 0$, and provided $1 + \xi (y^{(k)}_i-\mu)\sigma > 0$ for all $i$ and $k$, and
\begin{equation}
\ell(\mu,\sigma) = mr \log \sigma - \displaystyle\sum^m_{i=1}\exp\left\{-\left( \dfrac{y_i^{(r)}-\mu}{\sigma}\right)\right\}
- \displaystyle\sum^m_{i=1}\sum^r_{k=1} \dfrac{y_i^{(k)}-\mu}{\sigma},
\label{eq_rlos_like_0}
\end{equation}
if $\xi = 0$. Numerical methods can be implemented to find estimates of $\mu$, $\sigma$ and $\xi$. When $r=1$, equations (\ref{eq_rlos_like}) and (\ref{eq_rlos_like_0}) become the likelihood functions for the standard BM approach (equations (\ref{eq_like}) and (\ref{eq_like_0}), respectively). 

\subsection{Permutation bootstrapping}
\label{sec_boot}    

Let us consider a sequence of random variables ${\bf X} = X_1,\ldots,X_n$. This sequence is called exchangeable if, for all permutations $\Pi = \pi_1,\ldots,\pi_n$ of the integers $1,\ldots,n$, the sequence ${\bf X}_\Pi = X_{\pi_1},\ldots,X_{\pi_n}$ has the same joint distribution as ${\bf X}$. Further, if $\hat\theta = \hat\theta({\bf X})$ is some statistic computed from the sequence ${\bf X}$, we define the permutation bootstrapping (PB) statistic as
\begin{equation}
\hat\theta_{PB} = \dfrac{1}{n!}\sum_{\textrm{all }\Pi} \hat\theta({\bf X}_\Pi).
\label{eq_pb}
\end{equation}
\cite{Mefleh_2021} demonstrated that, if the sequence is exchangeable and $\hat\theta$ has finite variance, then the expected values of $\hat\theta_{PB}$ and $\hat\theta$ are equal, and the variance of $\hat\theta_{PB}$ is at most equal to that of $\hat\theta$. In practice, only approximate values of $\hat\theta_{PB}$ are calculated: because of the typical high values of $n!$, usually a small fraction of the permutations of $1,\ldots,n$, selected randomly, is taken into account when computing the PB statistic. As our observations consist of i.i.d. random variables, they are exchangeable, thus the PB approach is suitable to the task of estimating the GEV distribution parameters, whether with standard BM (as performed by \cite{Mefleh_2021}) or with $r$-LOS, as we propose in this work.

Throughout this study, we used $B=50$ permutations, given that \cite{Mefleh_2021} showed no substantial improvement in accuracy with higher values of $B$. Also, in all our calculations involving bootstrapping, we replaced the mean with the median in equation (\ref{eq_pb}), based on additional evidence by \cite{Mefleh_2021} indicating that the latter was more robust than the former to outliers.

\subsection{Simulations}
\label{sec_simul}

The procedure carried out to evaluate our new method was an adaptation of the one implemented by \cite{Mefleh_2021}. Our simulated time series consisted of sequences $\{ x_1,..., x_n \}$ $(n=365 \times 100)$ of independent samples from a Pareto random variable, i.e. they follow the cumulative distribution:
\begin{equation}
 \begin{array}{cc} 
F(x) = 1 - \left( \dfrac{1}{x} \right)^{1/\kappa}, & x > 1, 
\end{array}
\label{eq_pareto}
\end{equation}  
where $\kappa>0$. 
For each sequence, and following the new method, we computed 10 sets of estimates $\hat\mu$, $\hat\sigma$ and $\hat\xi$ of the GEV distribution parameters, one for each value of the order $r$. We also calculated, for each $r$, estimates $\hat{q}_{F,p}$ of the extreme quantile; we did this by replacing the values of $\hat\mu$, $\hat\sigma$ and $\hat\xi$ in equation (\ref{eq_quant}) with their estimates computed for each permutation and taking the median across all permutations. For a given set of constants (distribution parameter $\kappa$, order $r$, probability $p$), and using a constant block size $s=365$, we repeated this experiment $N' = 1000$ times, and calculated, across repetitions, the median absolute deviation (MAD) — a more robust alternative to the mean squared error metric, in terms of handling of outliers \citep{Leys_2013}, where one computes the median instead of the mean, and the absolute value instead of the square — between the true values of $\xi$ (or ${q}_{F,p}$) and their estimates computed with the new method; for the Pareto distribution, we used $\kappa$ as the true $\xi$. Additionally, we ran tests with data sampled not from a Pareto distribution, but from the Student's $t$ and the inverse gamma distributions -- the true values of the extreme value index for these distributions are, respectively, the inverse of the number of degrees of freedom \citep{Koedijk_1990, Huisman_2001, Schwaab_2021} and the inverse of the shape parameter \citep{Allouche_2024}.

\subsection{Real measurements}
\label{sec_real}

The real measurements we used for the evaluation of our method were the integer-valued maximum daily temperatures, in Fahrenheit, recorded in Fort Collins (CO, USA) between 1900 and 1999, and available with the R package \texttt{extRemes} \citep{Gilleland_2016}. Using block size $s=365$, we obtained estimates of $\xi$ and $q_{F,p}$ for $p = 1 - 1/(365\times 100)$, repeating the experiment $N' = 100$ times. In practice, this repetition of experiments simply consists in performing $50 \times N' $ permutations on the real data, and dividing these permutations into $N'$ groups for further analysis, because, unlike with our simulations, we worked with a single collection of real observations.

\section{Results}
\label{sec_results}

\subsection{Simulations}
\label{sec_res_sim}

Figures \ref{sim_block_g} and \ref{sim_block_q} show, respectively, MAD values for $\xi$ and $q_{F,p}$ (with $p = 1 - 1/n$) as a function of order $r$ for data sampled from a Pareto distribution (using 0.2, 0.5 or 0.8 for the distribution parameter $\kappa$); in order to highlight the impact of our method, these images also show estimates computed without permutations (i.e. standard BM and $r$-LOS parameter estimates). According to the plots in Figure \ref{sim_block_g}, the accuracy of the new method increased with at least some $r>1$ for all values of $\kappa$ analyzed; further, except for $\kappa = 0.2$, there was an order which resulted in an optimal parameter estimate ($r=9$ for $\kappa=0.5$ and $r=2$ for $\kappa=0.8$) — we ran tests with $\kappa=0.2$ and $r$ as high as 20 (data not shown), but still no optimal order was found. 
Finally, except when $\kappa=0.8$ (and only for higher values of $r$, which in any case did not correspond to cases with the best accuracy), the use of permutations was beneficial. The effects present in Figure \ref{sim_block_g} are also observed in Figure \ref{sim_block_q}, i.e. an improvement of performance with $r$, an optimum order when $\kappa=0.5$ or $\kappa=0.8$, and the positive impact of permutations. A different $p$ for the estimation of quantiles confirmed the previous findings, as seen in Figure \ref{sim_block_qb}. 

As in Figures \ref{sim_block_g} and \ref{sim_block_q}, Figure \ref{sim_block_t} displays MAD values for $\xi$ and $q_{F,p}$ as a function of order $r$ for data sampled from a Student's $t$ distribution with 5 degrees of freedom. We can see from these images that, as with the Pareto distribution (especially with higher $\kappa$), higher orders improved the estimation both of $\xi$ (up to $r=5$) and of $q_{F,p}$ (up to $r=9$), and MAD was lower when permutations are used. Finally, we ran tests with data sampled from an inverse gamma distribution with shape parameter 5 and scale parameter 1. As shown in Figure \ref{sim_block_ig}, the performance of the new method with this distribution was similar to that with the Pareto distribution, in particular when $\kappa=0.2$: values of MAD decreased with $r$, and the estimates were more accurate when permutations were used.

\subsection{Real measurements}
\label{sec_res_real}

\begin{figure}[b]
\centering
\includegraphics[width=0.4\textwidth]{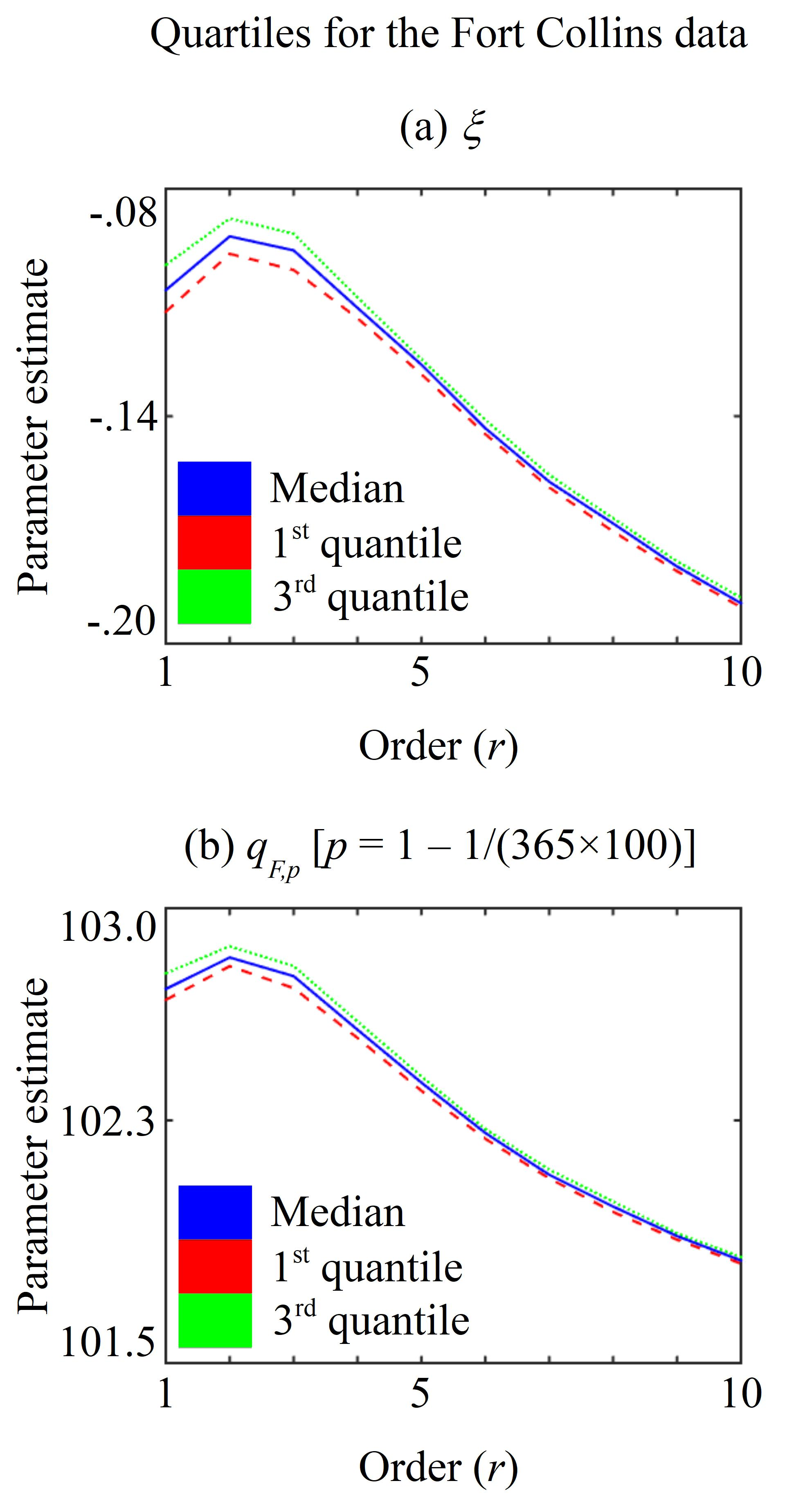}
\caption{Median (blue, solid lines), first quartile (red, dashed lines) and third quartile (green, dotted lines) of estimates of the extreme value index $\xi$ (plot (a)) and the quantile $q_{F,p}$, where $p=1-1/(365 \times 100)$ (plot (b)) as a function of order $r$, obtained from the Fort Collins maximum daily temperature data. }
\label{real}
\end{figure}

In Figure \ref{real}, we present plots for the estimates across all $N'=100$ experiment repetitions of the extreme value index and the $[1 - 1/(365\times 100)]^\textrm{th}$ quantile ($s=365$ being the block size), obtained with the method presented here, applied to the Fort Collins maximum daily temperature data; in these images, the lines represent the median estimates across all repetitions and also the first and third quantiles. The values that appear in these plots are consistent with the parameter and quantile estimates produced by \cite{Mefleh_2021} in their study with this data. Another noteworthy aspect of the effect of the new method on the meteorological time series is the reduced variability as $r$ increases, in both cases (extreme value index and quantile estimation).

\section{Discussion}
\label{sec_disc}

In this work, our main objective was to propose a new method to estimate parameters of the generalized extreme value distribution, one of the most popular mathematical tools in the field of extreme value theory. Our method combines the technique proposed by \cite{Mefleh_2021}, which applies bootstrapping to the standard block maxima framework, with the $r$ largest order statistics approach \citep{Smith_1986, Coles_2001, Zhang_2004}, or a generalization of block maxima whereby a number of the highest valued observations within a block of data, and not just the maxima, are taken into account for parameter estimation. As seen especially with simulated time series, our method had, in all cases analyzed, a performance that was better than that of either bootstrapping or $r$-LOS implemented individually. Thus, given the evidence presented here, we submit that our method is an attractive alternative for the estimation of GEV parameters. Concerning the results obtained with meteorological measurements, the values of the estimates computed with the new method were consistent with those calculated with permutation bootstrapping \citep{Mefleh_2021}; another noteworthy effect was that higher values of $r$ caused a decrease in variability.

\begin{figure}[b]
\centering
\includegraphics[width=0.9\textwidth]{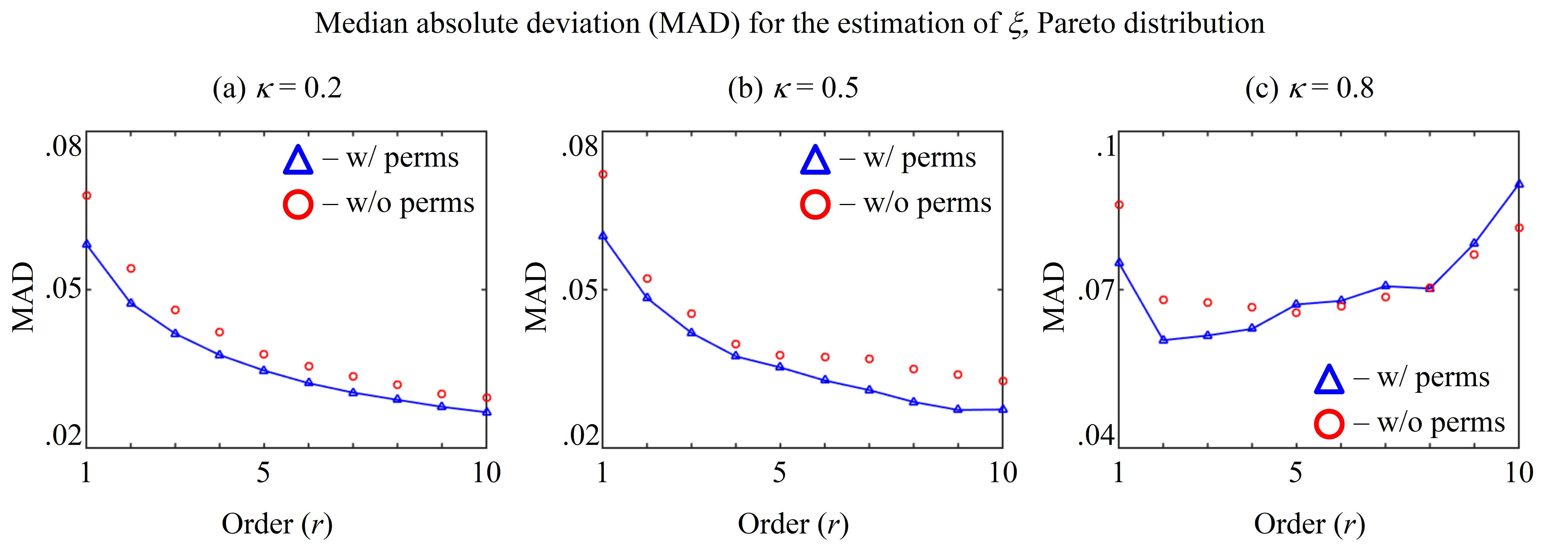}
\caption{Values of the median absolute deviation (MAD) for estimates of the extreme value index $\xi$ as a function of order $r$, for simulated data sampled from the Pareto distribution with three values of the distribution parameter: $\kappa=0.2$ (plot (a)), $\kappa=0.5$ (plot (b)) and $\kappa=0.8$ (plot (c)). In these plots, blue, connected triangles indicate the use of permutations; red, unconnected circles indicate that permutations were not used.  }
\label{sim_block_g}
\end{figure}

\begin{figure}[b]
\centering
\includegraphics[width=0.9\textwidth]{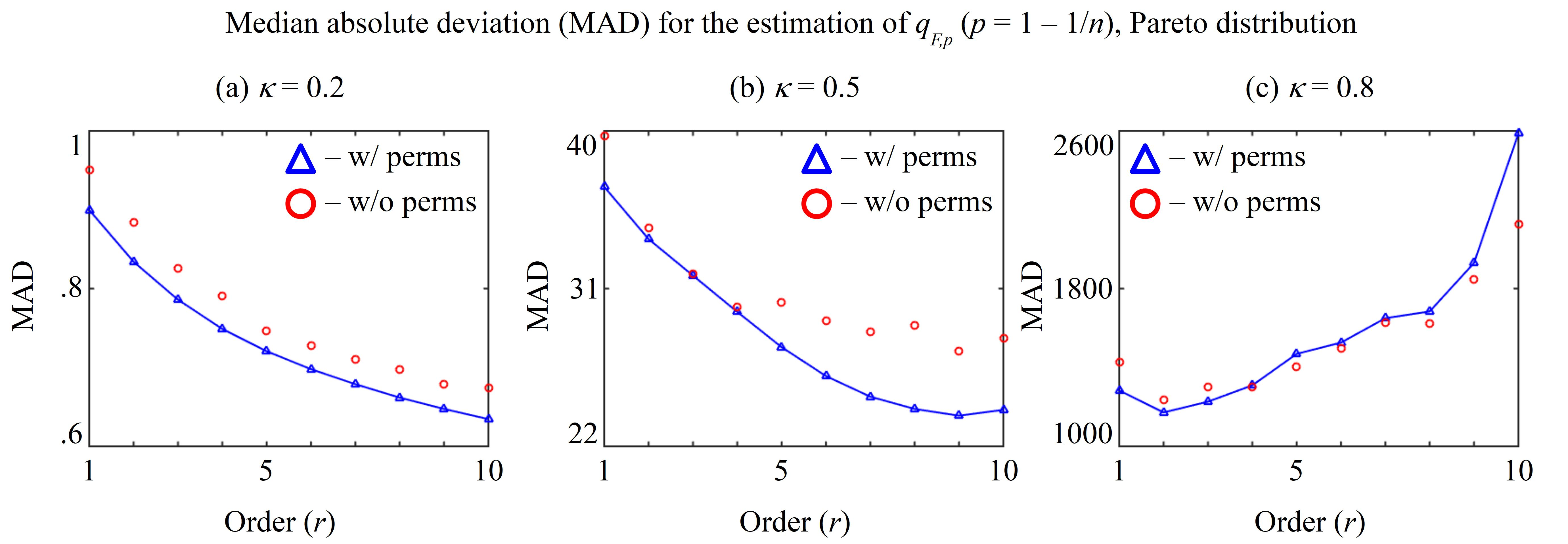}
\caption{Values of the median absolute deviation (MAD) for estimates of the quantile $q_{F,p}$ (where $p=1-1/n$ and $n=365 \times 100$) as a function of order $r$, for simulated data sampled from the Pareto distribution with three values of the distribution parameter: $\kappa=0.2$ (plot (a)), $\kappa=0.5$ (plot (b)) and $\kappa=0.8$ (plot (c)). The color coding and style for the lines of these plots was the same as in Figure \ref{sim_block_g}. }
\label{sim_block_q}
\end{figure}

\begin{figure}[b]
\centering
\includegraphics[width=0.4\textwidth]{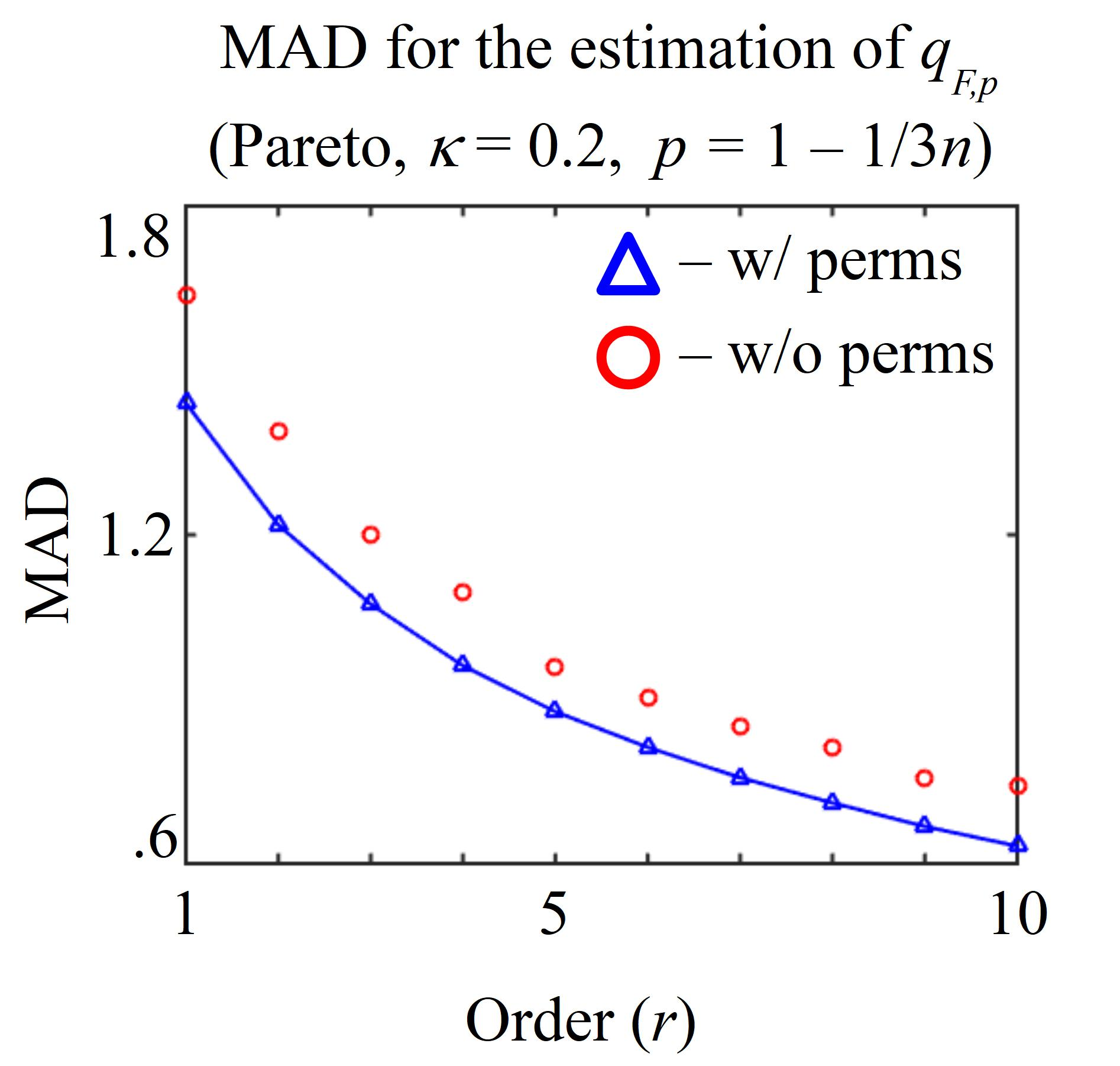}
\caption{Values of the median absolute deviation (MAD) for estimates of the quantile $q_{F,p}$ (where $p=1-1/3n$ and $n=365 \times 100$) as a function of order $r$, for simulated data sampled from the Pareto distribution with $\kappa=0.2$. The color coding and style for the lines of this plot was the same as in figure \ref{sim_block_g}. }
\label{sim_block_qb}
\end{figure}

\begin{figure}[b]
\centering
\includegraphics[width=0.4\textwidth]{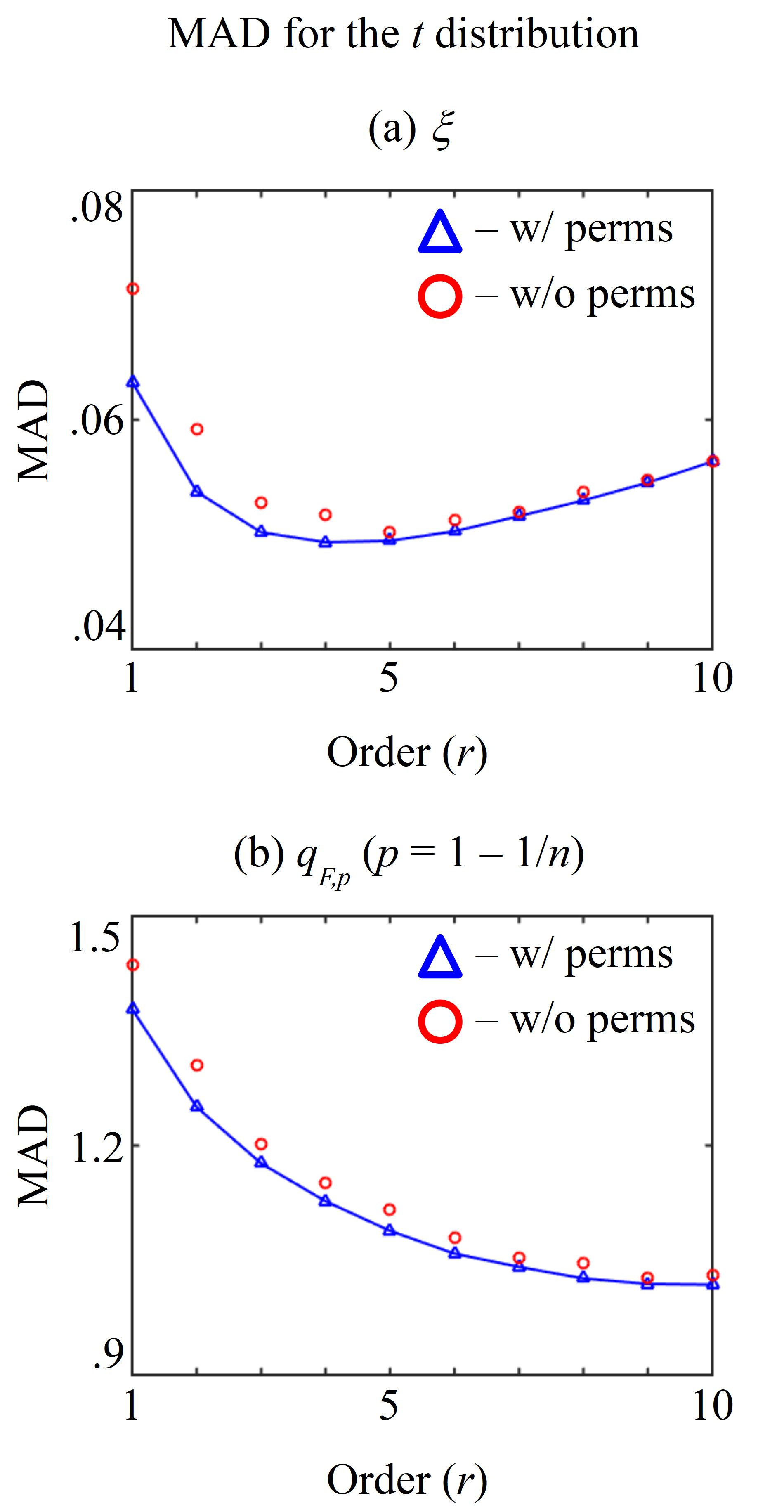}
\caption{Values of the median absolute deviation (MAD) for estimates of the extreme value index $\xi$ (plot (a)) and the quantile $q_{F,p}$ (where $p=1-1/n$ and $n=365 \times 100$) (plot (b)) as a function of order $r$, for simulated data sampled from Student's $t$ distribution. The color coding and style for the lines of these plots was the same as in figure \ref{sim_block_g}.}
\label{sim_block_t}
\end{figure}

\begin{figure}[b]
\centering
\includegraphics[width=0.4\textwidth]{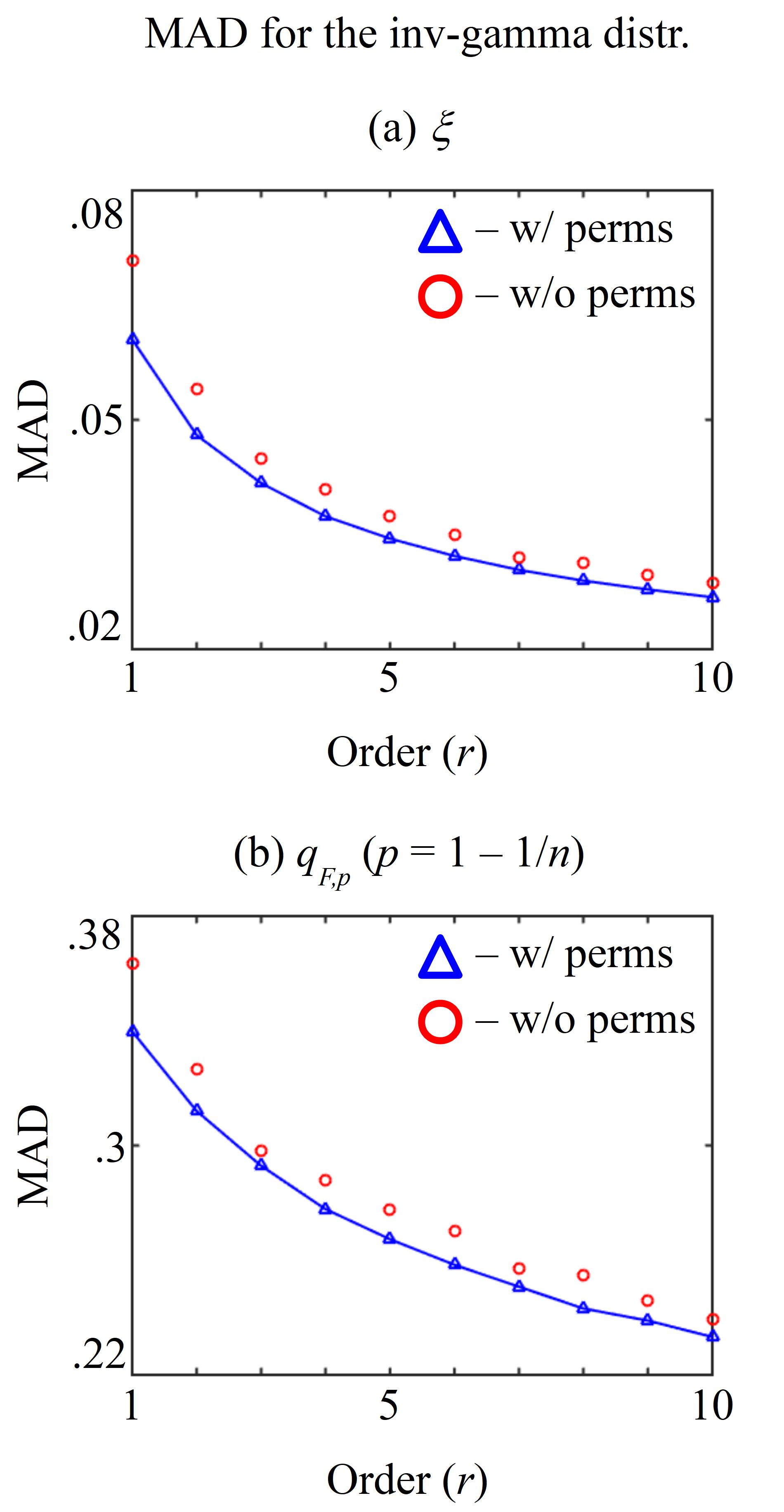}
\caption{Values of the median absolute deviation (MAD) for estimates of the extreme value index $\xi$ (plot (a)) and the quantile $q_{F,p}$ (where $p=1-1/n$ and $n=365 \times 100$) (plot (b)) as a function of order $r$, for simulated data sampled from the inverse gamma distribution. The color coding and style for the lines of these plots was the same as in figure \ref{sim_block_g}.}
\label{sim_block_ig}
\end{figure}

Based on the plots that appear in Figures \ref{sim_block_g} through \ref{sim_block_ig}, one could make the argument that our method of choice should be $r$-LOS instead of the new method, the reasons for this choice being: (a) $r$-LOS also outperforms standard BM and permutation bootstrapping; (b) the relative gain in accuracy provided by the new method over $r$-LOS is not substantial, as seen by the gaps between the triangles and the circles as a function of $r$ in all the plots; and (c) the new method requires much more computational resources, due to the additional calculations on the surrogate time series. We do not believe this to be the case. The advantages of using the new method instead of $r$-LOS (in the figures just mentioned, they can be seen when $r \neq 1$) are, in our view, comparable to those provided by the permutation bootstrapping approach proposed by \cite{Mefleh_2021} over standard BM (also in those same figures, now when $r=1$). Therefore, if the choice to implement permutation bootstrapping in lieu of standard BM is justified, which we think it is, then so is choosing the new method as the best of the three.

There were some important aspects related with parameter estimation for the GEV distribution that lay beyond the scope of this work and were therefore not dealt with here. One of them is the selection of $r$ for practical applications: as \cite{Coles_2001} and others have pointed out, there is a trade-off between bias and variance involved in this selection, smaller $r$ leading to higher variance, larger $r$ leading to bias. Automatic procedures to select $r$ have been proposed \citep{Bader_2018, Silva_2022}, while \cite{An_2007} provide simple rules for quick implementation; as the main purpose of this work was to demonstrate the validity and the better performance of the new method when compared with other approaches, we chose not to address this subject. Another issue we decided not to pursue was the impact of different dataset and block sizes, keeping constant $n=365 \times 100$ and $s=365$, respectively, both in simulation and real data calculations. With this, we attempted to mirror the procedures followed by many of the recent articles that apply the $r$-LOS framework (which our technique is based on) whose object of research is some natural phenomenon of a cyclical nature (yearly, say) for which there is abundant data, such as the weather \citep{Wang_2008, Sikhwari_2022} or sea levels \citep{Bader_2018, Naseef_2017, Haltas_2022}. 

Subsection \ref{sec_res_sim} showed probability distributions of different types varied their behavior as a function of $r$ and method of parameter estimation, i.e. the estimates for the Pareto data with higher $\kappa$ had some similarities with those computed from the $t$ distribution, while the findings with the inverse gamma distribution were closer to those obtained with the Pareto data with $\kappa=0.2$. While this behavior certainly merits a deeper investigation than that carried out in this article, that does not detract from the (in our view) more relevant conclusion that the new method was the more accurate of the three that we applied to time series with an array of distinct characteristics (e.g. Pareto and inverse gamma are one-tailed distributions whose support is a subset of the interval $(0,\infty)$, while the $t$ distribution is two-tailed and its support is the whole real line). Finally, we only looked into methods that dealt specifically with the GEV distribution, even though there have been some recent developments in the literature with promising results that are based on generalizations of that distribution, such as \cite{Shin_2023}. Nevertheless, it should be noted that, at least in principle, the framework proposed by \cite{Mefleh_2021} for BM and applied here to $r$-LOS can also be extended to methods such as the one presented by \cite{Shin_2023}.     

\bibliographystyle{plainnat}
\bibliography{draft_evt_subm_boot}

\end{document}